\documentclass{pnastwo}

\usepackage[dvips]{graphicx}
\usepackage{pnastwof}
\usepackage{amssymb,amsfonts,amsmath}
\usepackage{float}

\def\sles{\lower2pt\hbox{$\buildrel {\scriptstyle <}
   \over {\scriptstyle\sim}$}}

\def\sgreat{\lower2pt\hbox{$\buildrel {\scriptstyle >}
   \over {\scriptstyle\sim}$}}
\def\sharpnull#1{}

\def\apj{{\it Astrophys.~J.\ }}
\def\apjs{{\it Astrophys.~J. Suppl.\ }}

\def\mnras{{\it Mon.~Not.~R.~Astron.~Soc.\ }}

\url{www.pnas.org/cgi/doi/10.1073/pnas.0709640104}
\copyrightyear{2013}
\issuedate{Issue Date}
\volume{Volume}
\issuenumber{Issue Number}

\begin{document}

\title{The Astronomical Reach of Fundamental Physics}

\author{A. Burrows, J.P. Ostriker\affil{1}{Department of Astrophysical Sciences, Princeton University, Princeton, NJ\ 08544, USA}
}

\contributor{Accepted to Proceedings of the National Academy of Sciences
of the United States of America}

\maketitle

\begin{article}

\begin{abstract}

Using basic physical arguments, we derive by dimensional and physical analysis the characteristic masses
and sizes of important objects in the Universe in terms of just a few fundamental 
constants.  This exercise illustrates the unifying power of physics
and the profound connections between the small and the large in 
the Cosmos we inhabit. We focus on the minimum and maximum masses 
of normal stars, the corresponding quantities for neutron stars, the maximum 
mass of a rocky planet, the maximum mass of a white dwarf, 
and the mass of a typical galaxy.  To zeroth order, we show 
that all these masses can be expressed in terms of either the 
Planck mass or the Chandrasekar mass, in combination with various 
dimensionless quantities.  With these examples we expose the deep 
interrelationships imposed by Nature between disparate realms of the Universe and 
the amazing consequences of the unifying character of physical law.

\end{abstract}



\section{Fundamental Physical Constants from which to Build the Universe}
\label{sec:constants}

One of the profound insights of modern science in general, and of physics in particular,
is that not only is everything connected, but that 
everything is connected quantitatively. Another is that there are 
physical constants of Nature that by their units, constancy in space and time, and magnitude
encapsulate Nature's laws.  The constant speed of light delimits and defines 
the fundamental character of kinematics and dynamics.  Planck's constant
tells us that there is something special about angular momentum and the products of length and 
momentum and of energy and time.  What is more, in combination the constants of Nature
set the scales (lengths, times, and masses) for all objects 
and phenomena in the Universe, for scales are dimensioned entities and the only 
building blocks from which to construct them are the fundamental constants around 
which all Nature rotates.  Though in part merely dimensional analysis,
such constructions encapsulate profound insights into diverse physical phenomena.  

Hence, the masses of nuclei, atoms, stars, and galaxies are set by a restricted collection of basic
constants that embody the finite number of core natural laws.  In this paper, we 
demonstrate this reduction to fundamentals by deriving the characteristic masses 
of important astronomical objects in terms of just a few fundamental constants.
In doing so, we are less interested in precision than illumination,
and focus on the orders of magnitude.  Most of our arguments 
 are not original (see \cite{weiss,carr,rees,press,padman}, and references 
 therein), though some individual arguments are. This exercise will be valuable to the 
extent that it provides a unified discussion of the physical scales found 
in the astronomical world.  Those interested in the fundamental connections 
between the small and the large in this Universe we jointly inhabit are invited 
to contemplate the examples we have assembled here.

Reducing, even in approximate fashion, the properties of the objects of the Universe to 
their fundamental dependencies requires first and foremost a choice of irreducible 
fundamental constants. Various combinations of those constants of Nature can also be useful, 
so the word "irreducible" is used here with great liberty. We could choose among the following:
\begin{equation}
\label{constants}
\hbar, c, e, G, m_p, m_{\pi}, m_e, 
\end{equation}
where $\hbar$ is the reduced Planck's constant ($h/2\pi$), $c$ is the speed of light, 
$e$ is the elementary electron charge, $G$ is Newton's gravitational constant, 
$m_p$ is the proton mass, $m_{\pi}$ is the "pion" mass, and $m_e$ is the electron mass.  
%
%
It is in principle possible that these masses can be reduced to one fiducial mass, with
mass ratios derived from some overarching theory, but this is currently beyond the 
state of the art in particle physics. However, dimensionless combinations of fundamental 
constants emerge naturally in the variety of contexts
in which they are germane. Examples are $\alpha = e^2/\hbar{c}$, the fine structure
constant ($\sim$$\frac{1}{137}$); $\alpha_g = G m_p^2/\hbar{c}$ ($\sim$10$^{38}$), the corresponding gravitational coupling constant;
and $m_{pl} = ({\hbar{c}}/{G})^{1/2}$, the Planck mass ($\sim$$2\times 10^{-5}$ g).

As noted, if our fundamental theory were complete we would be able to express
all physical quantities in terms of only three dimensioned quantities, one each for 
mass, length, and time. Many would associate this fundamental theory with the Planck scale, 
so everything could be written in terms of $m_{pl}$, the Planck length ($R_{pl}$ = ($G{\hbar}/{c^3})^{1/2} \sim 10^{-33}$ cm), and the Planck 
time ($G{\hbar}/{c^5})^{1/2}$ ($\sim$$10^{-43}$ sec). The remaining quantities of relevance would be dimensionless
ratios derivable in this fundamental theory.  For instance, these ratios 
could be:

\begin{eqnarray}
\eta_p = m_{pl}/m_{p} (\sim 10^{19})\, , 
\nonumber\\
\eta_e = m_{pl}/m_e (\sim 10^{22})\, , 
\nonumber\\
\eta_{\pi} = m_{pl}/m_{\pi} (\sim 10^{20})\, , 
\end{eqnarray}
and $\alpha$, and, in principle, these ratios could be derived in the hypothetical fundamental theory.
We show in this paper that we can write astronomical masses in terms of $m_{pl}$ or $m_{p}$, with mass ratios and
dimensionless combinations of fundamental constants setting the corresponding relative
scale factors.  We thereby reduce "all" masses to combinations of only five quantities: $m_{p}$, $m_{e}$, $m_{\pi}$,
$\alpha$, and $\alpha_{g}$; or $m_{pl}$, $\eta_{p}$, $\eta_{e}$, $\eta_{\pi}$, and $\alpha$.  For radii, $\hbar$ 
and $c$ are explicitly needed. Note that $\alpha_g = 1/\eta^2_p$ and that $\alpha/\alpha_g \sim 10^{36}$.  
The latter is a rather large number, a fact with significant consequences.

We assume in this simplified treatment that the proton and neutron masses are the same, and equal to the atomic
mass unit, itself the reciprocal of Avogadro's constant ($N_A$).  The "pion" mass
can be the mass of any of the three pions, sets the length scale for the nucleus, and 
helps set the energy scale of nuclear binding energies \cite{clayton}.  However, for specificity, and 
for sanity's sake, we will assume that the number of spatial dimensions is three.  
Then, with only five constants we can proceed to explain, in broad outline, 
objects in the Universe.  In this essay, we emphasize stars (and planets) and
galaxies, but in principle life and the Universe are amenable to similar analyses \cite{rees_ostriker,silk,rees}.

\subsection{Nuclear Physical Scales}
\label{nuc_scales}

Before we proceed, we make an aside on nuclear scales and why we have introduced the pion mass and $\eta_{\pi}$.
Nucleons (protons and neutrons) are comprised of three quarks, and pions
are comprised of quark-antiquark pairs. Quantum Chromodynamics (QCD) \cite{greiner} is the 
fundamental theory of the quark and gluon interactions and determines the
properties, such as masses, of composite particles. The proton mass 
($\sim$938 MeV/c$^2$) is approximately equal to $\sim$$3\times \Lambda_{QCD}$, where $\Lambda_{QCD}$ ($\sim$300 MeV)
is the QCD energy scale \cite{wilczek}, and the "3" is its number of constituent 
quarks \cite{weinberg}.  The pion, however, is a pseudo-Goldstone boson and the square of its 
mass is proportional to $\Lambda_{QCD}(m_u + m_d)$, where $m_u$ and $m_d$
are the up and down bare quark masses, respectively.  $m_u + m_d$ is very approximately 
equal to 10 MeV (quite small). In principle,  all the hadron masses and 
physical dimensions can be determined from the bare quark masses and 
the QCD energy scale \cite{durr}, since the associated running coupling constant 
(which sets, for example, $g_{\pi{}NN}^2/4\pi$) perforce approaches "unity" (large 
values) on the very spatial scales that set particle size. 

For our purposes, we assume there are two fundamental masses in the theory.
These could be $\Lambda_{QCD}$ and $m_u + m_d$, but they could also be $m_p$ and $m_{\pi}$.
We opt for the latter.  Importantly, pion exchange mediates the strong force
between nucleons, and the nucleon-nucleon interaction determines nuclear binding energies
and nuclear energy yields.  Moreover, since the pion interaction is a derivative interaction \cite{weinberg},
the corresponding nuclear energy scale is not $m_{\pi}$ ($\sim$135 MeV/c$^2$), but is proportional to 
$\bigl(g_{\pi{}NN}^2/4\pi{}\bigr)m_{\pi}\bigl(\eta_{p}/\eta_{\pi}\bigr)^2$.  The upshot is that the binding energies of nuclei
scale as $m_{\pi}\bigl(\eta_{p}/\eta_{\pi}\bigr)^2$ and the size of the nucleon is $\sim$$\hbar/m_{\pi}c$,
about one fermi for the measured value of $m_{\pi}$.  Hence, the two "fundamental"
masses, $m_{\pi}$ and $m_{p}$, determine the density of the nucleus and nuclear 
binding energies, both useful quantities.  This fact is the reason we include 
$m_{\pi}$ (or $\eta_{\pi}$) in our list of fundamental constants with which we describe the Cosmos $-$
reasons for including $m_{p}$ (or $\eta_{p}$) are more self-evident.  An ultimate theory
would provide the ratios of all the couplings (QCD, electro-weak, gravitational)
and mass ratios, as well as all other dimensionless numbers of Nature.  Currently 
absent such a theory, we proceed using the quantities described.

\section{The Maximum Mass of a White Dwarf: The Chandrasekhar Mass}
\label{chandra}

Stars are objects in hydrostatic equilibrium for which inward forces
of gravity are balanced by outward forces due to pressure gradients.   
The equation of hydrostatic equilibrium is:

\begin{equation}
\label{HE}
\frac{dP}{dr} = -\rho \frac{G M(r)}{r^2}\, ,
\end{equation}
where $P$ is the pressure, $r$ is the radius, $\rho$ is the mass density, and $M(r)$
is the interior mass (the volume integral of $\rho$).  Dimensional analysis
thus yields for the average pressure or the central pressure ($P_c$) an approximate
relation, $P_c \sim \frac{G M^2}{R^4}$, where $M$ is the total stellar mass
and $R$ is the stellar radius and we have used the crude relation $\rho \sim \frac{M}{R^3}$.

For a given star, its equation of state, connnecting pressure with temperature, density, 
and composition, is also known independently.  Setting the central pressure derived using
hydrostatic equilibrium equal to the central pressure from thermodynamics or statistical physics
can yield a useful relation between $M$ and $R$ (for a given composition).  For example, if the 
pressure is a power-law function of density alone (i.e., a "polytrope"), such that $P = \kappa \rho^{\gamma}$, then 
using $\rho \sim \frac{M}{R^3}$ and setting $\kappa \rho^{\gamma}$ equal to $\frac{G M^2}{R^4}$
gives us 
\begin{equation}
\label{poly}
M \propto \Bigl(\frac{\kappa}{G}\Bigr)^{\frac{1}{2-\gamma}}R^{(4-3\gamma)/(2-\gamma)}\, .
\end{equation}
One notices immediately that $M$ and $R$
are decoupled for $\gamma = 4/3$. In fact, one can show from energy arguments that 
if $\gamma$ is the adiabatic gamma and not merely of structural significance, then 
at $\gamma = 4/3$ the star is neutrally stable $-$ changing its radius at a given mass 
requires no work or energy.  In other words, the star is unstable to collapse for $\gamma < 4/3$
and stable to perturbation and pulsation at $\gamma > 4/3$.  For $\gamma = 4/3$,
there is only one mass and it is $\bigl(\frac{\kappa}{G}\bigr)^{3/2}$.  

The significance of this is that white dwarf stars are supported by electron degeneracy 
pressure, fermionic zero-point motion, which is independent of temperature.
Such degenerate objects are created at the terminal stages
of the majority of stars and are what remains after such a star's thermonuclear life. 
The pressure, which like all pressures resembles an energy density, is approximately given 
by the average energy per electron (${<}\varepsilon{>}$) times the number density of electrons ($n_e$).  
If the electrons are non-relativistic, ${<}\varepsilon{>} \sim P_F^2/2m_e$, where $P_F$ is 
the fermi momentum, and simple quantum mechanical phase-space arguments for fermions give $P_F = (3\pi^2\hbar^3 n_e)^{1/3}$.
Since $n_e = N_A\rho Y_e$, where $Y_e$ is the number of electrons per baryon ($\sim$0.5), it is easy to show
that the pressure is proportional to $\rho^{5/3}$, that the associated $\gamma = 5/3$, 
and that the star is, therefore, a polytrope.  By the arguments above, such stars are stable.  

However, as the mass increases, the central density increases, thereby increasing the 
fermi momentum.  Above $c P_F \sim m_e c^2$, the electrons become relativistic.  
The formula for the fermi momentum is unchanged,
but the fermi energy is now linear (not quadratic) in $P_F$.  This fact means that ${<}\varepsilon{>}\times n_e$
is proportional to $\rho^{4/3}$, and that the white dwarf becomes unstable (formally 
neutrally stable, but slightly unstable if general-relativistic effects are included).
Hence, the onset of relativistic electron motion throughout a large fraction of the star, 
the importance of quantum mechanical degeneracy, and the inexorable effects of gravity conspire to yield 
a maximum mass for a white dwarf \cite{chandra,stoner}.  This mass, at the confluence of relativity ($c$), quantum mechanics ($\hbar$),
and gravitation ($G$), is the Chandrasekhar mass, named after one of its discoverers.  Its value can be derived
quite simply using $\bigl(\frac{\kappa}{G}\bigr)^{3/2}$ (eq. \ref{poly}) 
and $\kappa = \frac{3\pi^2}{4}\frac{\hbar{c}}{m_p^{4/3}}Y_e^{4/3}$, where we have set $N_A = 1/m_p$. 
The result is:
\begin{equation}
\label{eq:chandra}
M_{Ch} \sim \Bigl(\frac{\hbar{c}}{G}\Bigr)^{3/2}\frac{Y_e^2}{m_p^2}\, .
\end{equation}
Note that the electron mass does not occur and that the proper prefactor is $\sim$3.1.   
For $Y_e = 0.5$, $3.1Y_e^2$ is of order unity and $M_{Ch} = 1.46$ M$_{\odot}$.  
We have recovered our first significant result. There is a maximum mass for a white dwarf and its 
value depends only on fundamental constants $\hbar$, $c$, $G$, and $m_p$, most of 
which are of the microscopic world.  We can rewrite this result in numerous ways, some of which are:
\begin{eqnarray}
\label{eq.chandra2}
M_{Ch} \sim m_p \Bigl(\frac{\hbar{c}}{Gm_p^2}\Bigr)^{3/2}  = \frac{m_p}{\alpha_g^{3/2}}\, ,
\nonumber \\
M_{Ch} \sim m_{p} \eta^3_p \, , 
\nonumber \\
M_{Ch} \sim m_{pl} \eta^2_p \, .
\end{eqnarray}
Hence, the Chandrasekhar mass can be considered to be a function of the Planck mass 
and the ratio, $\eta_p$, or the proton mass and the small dimensionless gravitational 
coupling constant, $\alpha_g$ ($=\frac{Gm_p^2}{\hbar{c}}$). Such relationships are impressive 
in their compactness and in their implications for the power of science to explain Nature
using fundamental arguments.

\subsection{Radius of a White Dwarf}
\label{wdrad}

As an aside, we can use the arguments above to derive a relation
for the characteristic radius of a white dwarf. To do this we return to eq. (\ref{poly}), but 
insert $\gamma = 5/3$, the non-relativistic value relevant for most of the family.  The
degeneracy pressure is then given by $P = \kappa \rho^{5/3}$, where
\begin{equation}
\label{non-rel}
\kappa = \frac{(3\pi^2)^{2/3}}{5} \frac{\hbar^2}{m_e m_p^{5/3}} Y_e^{5/3}\, ,
\end{equation}  
an expression easily shown to follow from the general fact that pressure is 
average particle energy times number density, but for the non-relativistic relation
between momentum and energy.  We will find this relation useful elsewhere, so we highlight it.
Equation (\ref{poly}) then yields $R_{wd} \sim \frac{\kappa}{GM^{1/3}}$.  We now use some sleight-of-hand.
Noting that a white dwarf's mass changes "little" as the electron's fermi energy transitions
from non-relativistic to relativistic and the mass transitions from $\sim$0.5 M$_{\odot}$
to M$_{Ch}$, and using eq. (\ref{non-rel}), we obtain:
\begin{eqnarray}
\label{wdradius}
R_{wd} \sim \frac{\hbar}{m_e c} \eta_p  \, ,
\nonumber\\
\sim R_{pl} \eta_e \eta_p  \, ,
\nonumber\\
\sim 10^4 {\rm km}\, ,
\end{eqnarray}
where we have set $M \sim M_{Ch}$.
This relation powerfully connects the radius of a white dwarf to  
the product of the electron Compton wavelength and the ratio of the Planck mass to the proton mass.
It also provides the white dwarf radius in terms of the Planck radius.
Putting numbers in yields something of order a few thousand kilometers, the actual
measured value.  We will later compare this radius with that obtained for 
neutron stars, and derive a sweet result.

\section{Maximum Mass of a Neutron Star}
\label{maxNS}

Neutron stars are supported against gravity by pressure due to the strong repulsive nuclear 
force between nucleons.  Atomic nuclei are fragmented into their component subatomic particles and 
experience a phase transition to nucleons near and above the density
of the nucleus.  As we have argued, the interparticle spacing in the nucleus is
set by the range of the strong force, which is the reduced pion Compton wavelength, $\lambda_{\pi} = \hbar/m_{\pi}c$.
Therefore, since the repulsive nuclear force is so strong, the average density of a 
neutron star will be set (very approximately) by the density of the nucleus ($\rho_{nuc} \sim \frac{m_p}{4\pi\lambda_{\pi}^3/3}$).
Though nucleons are fermions, the maximum mass of a neutron star is not determined by the 
special relativistic physics that sets the Chandrasekhar mass \cite{lattimer}.  Rather, the maximum mass
of a neutron star is set by a general relativistic (GR) instability. GR gravity is stronger
than Newtonian gravity, and at high enough pressures, pressure itself becomes a source, along with mass,
to amplify gravity even more.  The result is that increasing pressure eventually becomes self-defeating 
when "trying" to support the more massive neutron stars and the object collapses.  Whereas the collapse
of a Chandrasekhar mass white dwarf leads to a neutron star, the collapse of a critical 
neutron star leads to a black hole.  

The critical mass of a neutron star is reached as its radius shrinks to near its 
Schwarzschild radius, $\frac{2GM}{c^2}$, say to within a factor of two (but call this 
factor $\beta_n$).  GR physics introduces no new fundamental constants and involves only  
$G$ and $c$. Therefore, the condition for GR instability is $R = \frac{2GM\beta_n}{c^2}$.
We can derive the corresponding mass by using the relation $\rho \sim \frac{3M}{4\pi{R^3}}$.  The result
is:
\begin{eqnarray}
\label{maxNSeq}
M_{NS} \sim \Bigl(\frac{\hbar{c}}{G}\Bigr)^{3/2}\frac{1}{m_p^2}\Bigl(\frac{\eta_{\pi}}{2\beta_n{\eta_{p}}}\Bigr)^{3/2}  \, ,
\nonumber \\
\sim M_{Ch}\Bigl(\frac{\eta_{\pi}}{2\beta_n{\eta_{p}}}\Bigr)^{3/2}  \, ,
\nonumber\\
\propto \frac{m_p}{\alpha_g^{3/2}}\Bigl(\frac{\eta_{\pi}}{2\beta_n{\eta_{p}}}\Bigr)^{3/2}  \, ,
\nonumber\\
\propto m_{pl} \eta^2_p \Bigl(\frac{\eta_{\pi}}{2\beta_n{\eta_{p}}}\Bigr)^{3/2} \, .
\end{eqnarray}
Dropping $\beta_n$ ($\sim O(1)$), $M_{NS}$ is also even more simply written 
as $M_{Ch}\Bigl(\frac{\eta_{\pi}}{{\eta_{p}}}\Bigr)^{3/2}$. 
The Planck mass and the Chandrasekhar mass appear again, but this time 
accompanied by the proton-pion mass ratio and unaccompanied by $3.1 Y_e^2$.
For $\beta_n \sim 2$, given the values of $m_p$ and $m_{\pi}$ with which we are familiar, 
$M_{NS}$ is actually close to $M_{Ch}$.  Details of the still-unknown nuclear equation of state, GR, 
and the associated density profiles yield values for $M_{NS}$ between $\sim$1.5 M$_{\odot}$ and $\sim$2.5 
M$_{\odot}$, where eventually we would need to distinguish between the "gravitational" mass 
and the "baryonic" mass. We note here that the fact that the two common types of degenerate stars $-$ endproducts
of stellar evolution $-$ have nearly the same mass derives in part from the 
rough similarity of the pion and proton masses. 

Of course, the pion mass, $m_{\pi}$, enters through its role in the nuclear density, as does $\hbar$ and $m_p$.  
$c$ enters through GR and through its role in setting the range of the Yukawa potential.
$G$ enters due the perennial combat of matter with gravity in stars.  All but $m_{\pi}$ are 
actors in the Chandrasekhar saga, so we shouldn't be surprised 
to see $M_{Ch}$ once again, albeit with a modifying factor. Note that, since 
neutron stars are formed when a white dwarf achieves the Chandrasekhar mass,
and critical neutron stars collapse to black holes, the existence of stable neutron stars
requires that $M_{NS} > M_{Ch}$, with many, many details swept under the rug.  
When those details are accounted for we find that $M_{NS}$ (baryonic) is 
about twice $M_{Ch}$, the ratio being comfortably greater than, but at the same time uncomfortably 
close to, unity. Suffice it to say, stable neutron stars do exist in our Universe.

\subsection{Radius of a Neutron Star}
\label{NS-rad}

The radius of a neutron star is set by some multiple of the Schwarzschild radius.  
The baryonic mass of a neutron star is set by its formation and accretion history, but it is 
bounded by the minimum possible Chandrasekhar mass (recall this is set by $Y_e$).  
So, if we set a neutron star's mass equal to its maximum, we will not be far off.
The upshot is the formula:
\begin{equation}
\label{eq:NSrad}
R_{ns} \sim 2\frac{2G}{c^2}\Bigl(\frac{\hbar{c}}{G}\Bigr)^{3/2}\frac{1}{m_p^2}
\nonumber \\
\sim \frac{\hbar}{m_p{c}}  \eta_p
\nonumber \\
\sim R_{pl} \eta^2_p \, . 
\end{equation}  
This equation is very much like eq. (\ref{wdradius}) for the radius of a white dwarf, with $m_p$ substituted
for $m_e$.  Therefore, the ratio of the radius of a white dwarf to the radius of a neutron star
is simply the ratio of the proton mass to the electron mass ($\eta_e/\eta_p$), within factors of order unity.  
In our Universe, $\eta_e/\eta_p$ is $\sim$1836, but we can call this "three orders of magnitude."  Ten-kilometer neutron stars
quite naturally imply $\sim$$5\times 10^3$- to $\sim$10$^4$-kilometer white dwarfs.  By now, 
we shouldn't be surprised that this ratio arises in elegant fashion from fundamental 
quantities using basic physical arguments.

\section{Minimum Mass of a Neutron Star}

The minimum possible mass ($M_{ns}$) for a stable neutron star may not be easily 
realized in Nature. Current theory puts it at $\sim$0.1 M$_{\odot}$ \cite{haensel}, and there is no realistic mechanism
by which a white dwarf progenitor near such a mass, inside a star in its terminal stages or in a 
tight binary, can be induced to collapse to neutron-star densities. The Chandrasekhar
instability is the natural agency, but $M_{Ch}$ is much larger than $\sim$0.1 M$_{\odot}$. 
However, such an object might form via gravitational instability
in the accretion disks of rapidly rotating neutron stars of canonical mass. 
Mass loss after the collapse of a Chandrasekhar core is possible by Roche-lobe overflow
in the tight binary, but the associated tidal potential is quite different from the
spherical potential for which $M_{ns}$ is derived. 

Nevertheless, a first-principles estimate of $M_{ns}$ is of some interest.   What is 
the physics that determines it?  A white dwarf is supported by the electron degeneracy pressure
of free electrons, and its baryons are sequestered in nuclei.  A neutron star is supported
by the repulsive strong force between degenerate free nucleons, and most
of its nuclei are dissociated.  Both are gravitationally bound.  One can ask the question: 
for a given mass, which of the two, neutron star or white dwarf, is the lower energy state?
Note that to transition to a neutron star, the nuclei of a white dwarf must be dissociated
into nucleons, and the binding energy of the nucleus must be paid.  Note also that a neutron
star, with its much smaller radius, is the more gravitationally bound.  Therefore, we see that when the 
specific (per nucleon) gravitational binding energy is equal to the specific nuclear binding energy
of the individual nuclei, we are at $M_{ns}$. Above this mass, the conversion
of an extended white dwarf into a compact neutron star and the concomitant release of 
gravitational binding energy can then pay the necessary nuclear breakup penalty. 
But, of course, a substantial potential barrier must be overcome.

In equation form, this can be stated as $\frac{GM^2}{R} \sim \Delta{M}c^2$, where $\Delta{M}c^2$ is the binding
energy of the nucleus.  This equation can be rewritten: $\frac{GM}{Rc^2} \sim \frac{\Delta{M}}{M} = f$, where $f$ 
is the binding energy per mass, divided by $c^2$. From the fact that 
the nucleon is bound in a nucleus with an energy $\sim$$m_{\pi}c^2(\frac{\eta_{p}}{\eta_{\pi}})^2$, we obtain
$f \sim (\frac{\eta_{p}}{\eta_{\pi}})^{3}$ (an appropriate extra factor might be $\sim$3).  
In our Universe, $f$ is $\sim$0.01$-$0.02.  Now, notice that we can rewrite this condition as: 
$\frac{2GM}{Rc^2} \sim 2f$, which is the same as the condition for the maximum neutron star mass, with
$\frac{1}{2f}$ substituted for $\beta_n$ in eq. (\ref{maxNSeq}).  Using exactly the same manipulations, 
we then derive:
\begin{eqnarray}
\label{minNSmass}
M_{ns} \sim M_{Ch} f^{3/2} \Bigl(\frac{\eta_{\pi}}{\eta_{p}}\Bigr)^{3/2}  \, ,
\nonumber \\
\sim M_{Ch} \Bigl(\frac{\eta_{p}}{\eta_{\pi}}\Bigr)^{3}\, ,
\nonumber \\
\sim m_{pl} \eta^2_p \Bigl(\frac{\eta_{p}}{\eta_{\pi}}\Bigr)^{3}\, .
\end{eqnarray} 
The Planck mass and $M_{Ch}$ pop up again (quite naturally), but this time accompanied by the small 
factor $(\frac{\eta_{p}}{\eta_{\pi}})^{3}$ ($(\frac{m_\pi}{m_p})^{3}$).  
Note that $M_{ns}/M_{NS} \sim (2\beta_n{f})^{3/2} \propto (\frac{m_{\pi}}{m_p})^{4.5}$,
fortunately a number less than one.  However, to derive the precise value for $M_{ns}$ requires 
retaining dropped factors and much more precision, but the basic dependence on  
fundamental constants is clear and revealing.

\section{Maximum Mass of a Rocky Planet}

A rocky planet such as the Earth, Venus, Mars, or Mercury is comprised of silicates 
and/or iron and is in hydrostatic equilibrium.  If its composition were uniform,  
its density would be constant at the solid's laboratory value.  This value is set by the Coulomb interaction and 
quantum mechanics (which establishes a characteristic radius, the Bohr radius).  
As its mass increases, the interior pressures increase and the planet's matter
is compressed, at first barely and slowly (because of the strength of materials), 
but later (at higher masses) substantially. Eventually, for the highest masses, many of the electrons in the 
constituent atoms would be released into the conduction band, the material would be 
metalized \cite{wigner}, and the supporting pressure would be due to degenerate electrons.
Objects supported by degenerate electrons are white dwarfs.  Therefore, at low masses
and pressures, the (constant) density of materials and the Coulomb interaction imply 
$R \propto M^{1/3}$, while at high masses the object acts like a white dwarf
supported by electron-degeneracy pressure and $R \propto M^{-1/3}$.
This bahavior indicates that there is a mass at which the radius is a maximum for a given composition \cite{zapolsky}.
It also implies that there is a mass range over which the radius of the object is
independent of mass and $\frac{dR}{dM} \sim 0$.  For hydrogen-dominated objects,
this state, for which Coulomb and degeneracy effects "balance," is where Jupiter and Saturn reside,
but one can contemplate the same phenomenon for rocky and iron planets.  It
is interesting to ask the question: what is the maximum mass ($M_{rock}$) of a "rocky"
planet, above which its radius decreases with increasing mass in white-dwarf-like fashion,
and below which it behaves more like a member of a constant-density class of objects?
This definition provides a reasonable upper bound to the mass of a rocky or solid planet.
With the anticipated discovery of many exoplanet "Super-Earths" in coming years, this
issue is of more than passing interest. 

There are two related methods to derive $M_{rock}$.  The first is to set the 
expression for non-relativistic degeneracy pressure ($P_0 = \kappa \rho_0^{5/3}$, where $\kappa$ is given 
by eq. \ref{non-rel}) equal to the corresponding expression for the 
central pressure of a constant density object in hydrostatic equilibrium.  The result is:
\begin{equation}
\label{conden}
P = \frac{1}{2}\Bigl(\frac{4\pi}{3}\Bigr)^{1/3} G M^{2/3} \rho_0^{4/3}\, .
\end{equation}
The other method is to use the radius-mass relations for both white dwarfs and constant-density
planets and to set the radius obtained using one relation equal to that using the other.  Stated
in equation form, this is $R_{wd} \sim \frac{\kappa}{GM^{1/3}} \sim R_{rock} \sim \bigl(\frac{3M}{4\pi\rho_0}\bigr)^{1/3}$.
For both methods one needs the density, $\rho_0$, and using either method 
one gets almost exactly the same result. $\rho_0$ is set by the Bohr radius, $a_B = \frac{\hbar^2}{m_e e^2}$,
and one then obtains:
\begin{equation}
\rho_0 \sim \frac{3 m_p\mu}{4\pi a_B^3}\, ,
\end{equation}
where $\mu$ is the mean molecular weight of the constituent silicate or iron (divided by $m_p$).
A bit of manipulation yields 
\begin{eqnarray}
\label{rock}
M_{rock} \sim m_p \Bigl(\frac{e^2}{Gm_p^2}\Bigr)^{3/2} \frac{6}{\pi} Y_e^{5/2}\mu^{1/3}  \, ,
\nonumber \\
\sim m_p \Bigl(\frac{\alpha}{\alpha_g}\Bigr)^{3/2}  \, ,
\nonumber \\
\sim M_{Ch} \alpha^{3/2} \, ,
\nonumber \\
\sim m_{pl} \eta^2_p \alpha^{3/2} \, .
\end{eqnarray}
Note that $M_{rock}$ is actually independent of $\hbar$, $c$, and $m_{e}$.  

This eqaution expresses another extraordinary result.  It indicates that, like 
other masses we analyze in this paper, $M_{rock}$ can be 
written to scale with the Planck mass and $\eta_p$ or the Chandrasekhar mass, but multiplied
by a power of the fine-structure constant.  The latter comes from the role of the 
Coulomb interaction in setting the size scale of atoms and also indicates that $M_{rock}$
is much smaller than $M_{Ch}$, as one might expect.  As eq. (\ref{rock}) also shows,
$M_{rock}$ scales with $m_p$, amplified by the ratio of the electromagnetic to the gravitational
fine-structure constants to the 3/2 power.  Plugging numbers into either of these formulae,
one finds that $M_{rock}$ is $\sim$2 $\times 10^{30}$ grams $\sim$300 $M_{Earth}$ $\sim$1 $M_{Jup}$.
Detailed calculations only marginally improve upon this estimate.
Note that reinstating the $Y_e^{5/2}\mu^{1/3}$ dependence yields roughly the same value 
for both hydrogen and iron planets.

\section{The Maximum Mass of a Star}
\label{maxstar}

The fractional contribution of radiation pressure ($P_{rad} = {(1/3)}a_{rad} T^4$,
where $a_{rad} = \frac{\pi^2k_B^4}{15\hbar^3{c^3}}$ and $k_B$ is 
Boltzmann's constant) to the total supporting pressure in stars 
increases with mass.  At sufficiently large mass, the gas becomes 
radiation-dominated and $P_{rad}$ exceeds the ideal gas contribution, 
$P_{IG} = \frac{\rho{k_B T}}{\mu{m_p}}$.  
The entropy per baryon of the star also rises with mass.  
If we set $\beta P \equiv P_{IG}$, where $P = P_{IG} + P_{rad}$, thereby defining $\beta$,
we derive \cite{eddington}:
\begin{equation}
\label{eddington}
P = \Bigl(\frac{1-\beta}{\beta^4}\Bigr)^{1/3}\Bigl(\frac{3k_B}{\mu{m_p}a_{rad}}\Bigr)^{1/3} \frac{k_B}{\mu{m_p}} \rho^{4/3}\, .
\end{equation}
The upshot is that the effective "polytropic" gamma (defined assuming $P \sim \kappa \rho^{\gamma}$) 
decreases from $\sim$5/3 towards $\sim$4/3 as the stellar mass increases. As we argued 
in the section on the Chandrasekhar mass, the onset of relativity makes a star susceptible to 
gravitational instability.  Photons are relativistic particles.  A radiation-pressure 
dominated stellar envelope can "easily" be ejected if perturbed, and at the very 
least is prone to pulsation if coaxed.  Such coaxing and/or perturbation
could come from nuclear burning or radiation-driven winds. With radiation domination, 
and given the high opacity of envelopes sporting heavy elements with high $Z$,   
radiation-pressure-driven winds can blow matter away and in this manner limit 
the mass that can accumulate.  This physics sets the maximum mass ($M_{S}$) of a stable star
on the hydrogen-burning "main sequence."  

Using eq. (\ref{poly}), with $\gamma = 4/3$, and eq. (\ref{eddington}),
and the same arguments by which we derived the Chandrasekhar mass, we
find the only mass for a given $\kappa$:
\begin{eqnarray}
\label{eq:maxstar}
M_{S} \sim \Bigl(\frac{\kappa_{rad}}{G}\Bigr)^{3/2}  \, ,
\nonumber \\
\sim \Bigl(\frac{1-\beta}{\beta^4}\Bigr)^{1/2} \Bigl(\frac{45}{\pi^2}\Bigr)^{1/2} \Bigl(\frac{\hbar{c}}{G}\Bigr)^{3/2}\frac{1}{m_p^2\mu^2}  \, ,
\nonumber \\
\sim \Bigl(\frac{1-\beta}{\beta^4}\Bigr)^{1/2} \Bigl(\frac{45}{\pi^2}\Bigr)^{1/2} \frac{M_{Ch}}{\mu^2}\, ,
\nonumber \\
\sim \Bigl(\frac{1-\beta}{\beta^4}\Bigr)^{1/2} \Bigl(\frac{45}{\pi^2}\Bigr)^{1/2}  m_{pl} \bigl(\eta_p/\mu\bigr)^{2}\, ,
\end{eqnarray}
where $M_{Ch}$ is our $Y_e$-free Chandrasekhar mass. Since $k_B$ has no 
meaning independent of temperature scale, it does not appear in eq. (\ref{eq:maxstar}).   
That $M_{Ch}$ occurs in $M_{S}$ should not be surprising, since in both cases
the onset of relativity in the hydrostatic context, for electrons in one case 
and via photons in the other, is the salient aspect of the respective limits \cite{adams}. 
In determining both $M_{Ch}$ and $M_{S}$, relativity ($c$), quantum 
mechanics ($\hbar$), and gravity ($G$) play central roles.

The ratio of $M_{S}$ to $M_{Ch}$($Y_e$) depends only on $\mu$, $Y_e$, $\beta$, 
and some dimensionless constants.  Therefore, it is a universal number.
For $\beta = 1/2$, $Y_e = 0.5$, and $\mu \sim 1$, we find  $M_{S}$/$M_{Ch}$ $\sim$ 20, but for smaller $\beta$s
(a greater degree of radiation domination), the ratio is larger ($\propto \frac{1}{\beta^2}$).  The relevant $\beta$ 
and actual ratio depend upon the details of formation and wind mass loss (and, hence, heavy element abundance),
and are quite uncertain.  Nevertheless, indications are that the latter ratio could range 
from $\sim$50 to $\sim$150 \cite{hump}.  

As an aside, we note that one can derive another (larger) maximum mass associated with general relativity  
by using the fact that the fundamental mode for spherical stellar pulsation occurs at a critical thermodynamic
gamma ($\gamma_1$)\footnote{$\gamma_1$ is the logarithmic derivative of the pressure with respect 
to the mass density at constant entropy.}.  Including the destabilizing effect of general relativity,
this critical $\gamma_1$ equals 4/3 + $K \frac{GM}{Rc^2}$, where $K \sim 1$.  A mixture of
radiation with ideal gas for which radiation is dominant has a $\gamma_1$ of $\sim$4/3 + $\beta /6$ \cite{clayton}.   
One could argue that, if the temperature is high enough to produce electron-positron
pairs, then $\gamma_1$ would plummet and the star would be unstable to collapse.
Hydrostatic equilibrium suggests that the stellar temperature times Boltzmann's constant 
when pairs start to become important would then be some fraction of $\sim$$m_e c^2$ 
and would equal $\frac{GMm_p}{R}$.  This correspondence gives us an expression for $\frac{GM}{Rc^2}$ ($\sim \eta_p/\eta_e$) and, 
therefore, that  $\beta /6 \sim m_e/m_p = \eta_p/\eta_e$. Using eq. (\ref{eq:maxstar}), we 
find a maximum stellar mass due to general-relativistic instability and pair production 
near $\sim$$10^6$ M$_{\odot}$, close to the expected value \cite{hoyle}. 
Since other instabilities intervene before this mass is reached, this limit 
is not relevant for the main-sequence limit.

However, the facts that 1) $M_{S}$ is much greater than $M_{Ch}$ and that 2) their ratio is 
a large constant is interesting.  Therefore, we can take some 
comfort in noting that since $M_{S}$ is much larger than $M_{Ch}$, white dwarfs, 
neutron stars, and stellar-mass black holes, all stellar "residues" 
that would be birthed in stars, are allowed to exist.

\section{Minimum Mass of a Star}

Stars are assembled from interstellar medium gas by gravitational collapse.
As they contract, they radiate thermal energy and the compact protostar that first
emerges is in quasi-hydrostatic equilibrium.  Before achieving the hydrogen-burning 
main sequence, the protostar becomes opaque and radiates from a newly-formed 
photosphere at a secularly evolving luminosity. The progressive loss of energy
from its surface occasions further gradual quasi-hydrostatic contraction.
In parallel, the  central temperature increases via what is referred to as the
"negative specific heat" effect in stars.  Energy loss leads to 
temperature increase.  From energy conservation, the change in gravitational 
energy (loosely $-\frac{GM^2}{2R}$) due to the shrinkage is equal to the sum of the photon losses 
and the increase in thermal energy, in rough equipartition. Using either hydrostatic
equilibrium or the Virial theorem, and the ideal gas law connecting pressure with temperature 
($P = \frac{\rho k_B T}{\mu m_p}$), we obtain
\begin{equation}
\label{temp}
k_B T_{int} \sim \frac{GMm_p\mu}{3R} \sim G\mu m_p M^{2/3} \rho^{1/3}\, ,
\end{equation} 
where $\mu$ is the "mean molecular weight" (of order unity), and $R \sim \bigl(\frac{3M}{4\pi \rho}\bigr)^{1/3}$ has been used.  
We note that eq. (\ref{temp}) demonstrates that the temperature increases 
as the star shrinks.  

However, the "negative specific heat" is an indirect consequence of the ideal gas law.
As the protostar contracts and heats, its density rises.  In doing so, 
the core entropy decreases (despite the temperature increase) and the core becomes
progressively more electron-degenerate.  When it becomes degenerate, because
degeneracy pressure is asymptotically independent of temperature,  
further energy loss does not lead to temperature increase, but decrease.
Hence, there is a peak in the core temperature that manifests itself at the onset 
of core degeneracy.  The standard argument states that if at this peak temperature
($T_{max}$) the integrated core thermonuclear power is smaller
than the surface luminosity (also a power), then the "star" will not achieve
the hydrogen main sequence.  It will become a "brown dwarf," which will cool
inexorably into obscurity, but over Gigayear timescales.  The mass at which $T_{max}$ is just sufficent
for core power to balance surface losses is the minimum stellar mass ($M_{s}$).
Below this mass is the realm of the brown dwarf. Above it reside canonical, 
stably-burning stars \cite{bur93}.                                                  

We can use eq. (\ref{temp}) to derive the critical mass in terms of $T_{max}$.
A gas becomes degenerate when quantum statistics emerges to be important.
For an electron, this is when the deBroglie wavelength of the electron 
($\lambda = \frac{h}{m_e v}$, where $v$ is the average particle speed)
approaches the interparticle spacing, $(\frac{\mu m_p}{\rho})^{1/3}$. 
It is also when the simple expression for the ideal gas pressure ($=\frac{\rho k_B T}{\mu m_p}$) 
equals the simple expression for the degeneracy pressure ($P = \kappa \rho^{5/3}$, where 
$\kappa$ is given by eq. \ref{non-rel}).  We use the latter condition
and derive an expression for $T_{max}$:
\begin{equation}
\label{temp2}
k_B T_{max} \sim \frac{\mu m_p}{\kappa} G^2 M^{4/3} \sim \frac{G^2 \mu m_p^{8/3} m_e}{\hbar^2 Y_e^{5/3}} M^{4/3}\, .
\end{equation}

The textbooks state that $M_s$ is then derived by setting $T_{max}$ to some 
"ignition" temperature (frequently set to $10^6$ K), and then solving for 
$M$ in eq. (\ref{temp2}). In this way, using measured constants and retaining a few dropped coefficients, 
one derives $M_s \sim 0.1$ M$_{\odot}$ and this number is rather accurate.  

However, this procedure leaves unexplained the origin of $T_{max} \sim 10^6$ K \footnote{a better 
number for $T_{max}$ is $\sim$$3.5\times 10^6$ K.} and here we depart from the traditional 
explanation to introduce our own. There are two things to note.  First, $M_s$
is not determined solely by thermonuclear considerations $-$ photon opacities, temperatures, densities, 
and elemental abundances ("metallicity") at the stellar surface set the emergent luminosity that 
is to be balanced by core thermonuclear power.  Second, the "ignition" temperature is not some fundamental quantity, 
but is in part determined by the specific nuclear physics of the relevant thermonuclear process, 
in this case the low-temperature exothermic proton-proton reactions to deuterium, $^3$He, 
and $^4$He.  Hence, we would need the details of the interaction physics of the proton-proton
chain, integrated over the Maxwell-Boltzmann distribution of the relative proton-proton
energies and over the temperature and density profiles of the stellar core. However,
a simpler approach is possible \cite{adams}.  We can do this because the thermonuclear 
interaction rate is greatest at large particle kinetic energies,
which are Boltzmann suppressed ($\propto e^{-{E}/{k_B T}}$), and because Coulomb
repulsion between the protons requires that they barrier-penetrate (quantum tunnel) to within range 
of the nuclear force.  This fact introduces the "Gamow" factor \cite{gamow}:
\begin{equation}
\label{gamow}
e^{-\frac{2e^2}{\hbar v}} = e^{-\frac{2\alpha c}{v}} = e^{-\frac{b}{\sqrt{E}}}\, ,
\end{equation}
where $E = 1/2 \mu_p v^2$, $\mu_p$ here is the reduced proton mass ($m_p/2$), 
$v$ is the relative speed, and this equation defines $b$ as $\alpha c \sqrt{\mu_p/2}$.
The reaction rate contains the product of the Boltzmann and Gamow exponentials and the necessary integral
over the thermal Maxwell-Boltzmann distribution yields another exponential of a function of $T$.
That function is determined by the method of steepest descent, whereby the product of the Boltzmann and Gamow factors
is approximated by the exponential of the extremal value of the argument: $-\frac{E}{k_B T} - \frac{b}{\sqrt{E}}$. 
That "Gamow peak" energy ($E_{gam}$) is $\bigl({b k_B T}/{2}\bigr)^{2/3}$. The result is
a term:
\begin{equation}
\label{gamow2}
e^{-\frac{E}{k_B T}}e^{-\frac{b}{\sqrt{E}}} \rightarrow e^{-\frac{3 E_{gam}}{k_B T}}
\end{equation}
in the thermonuclear rate.  Though the surface 
luminosity depends upon metallicity and details of the opacity, that dependence is not very 
strong \cite{bur93}.  Furthermore, due to the Boltzmann and Gamow exponentials (eq. 
\ref{gamow2}), any characteristic temperature we might derive (which might be 
set equal to $T_{max}$) depends only {\it logarithmically} on the problematic 
rate prefactor.  This argument allows us to focus on the argument of the exponential 
in eq. (\ref{gamow2}).  Therefore, we set $\frac{3 E_{gam}}{k_B T}$ in eq. (\ref{gamow2})
equal to a dimensionless number $f_g$ of order unity (but larger) and presume that
it, whatever its value, is a universal dimensionless number, to within logarithmic terms.
The resulting equation for the critical $T$ is:
\begin{equation}
\label{tmax}
k_B T_{max} = \frac{27}{16f_g^3} \alpha^2 m_p c^2\, .
\end{equation} 
In eq. (\ref{tmax}), both $\alpha$ and $m_p$ come from the 
$\frac{2e^2}{\hbar v}$ term associated with Coulomb repulsion in Gamow 
tunneling physics.
Setting $T_{max}$ in eq. (\ref{tmax}) equal
to $T_{max}$ in eq. (\ref{temp2}), we obtain:
\begin{eqnarray}
\label{mgamow}
M_s \sim m_{pl} \Bigl(\frac{\eta_{p}}{f_g^{9/8}}\Bigr)^2 \Bigl(\frac{\eta_e}{\eta_p} \alpha^2\Bigr)^{3/4} 
\Bigl(\frac{27\pi^2}{8}\Bigr)^{3/4} \frac{Y_e^{5/4}}{\mu^{3/4}}  \, ,
\nonumber \\
\sim \frac{M_{Ch}}{f_g^{9/4}} \Bigl(\frac{\eta_e}{\eta_p} \alpha^2\Bigr)^{3/4}
\Bigl(\frac{27\pi^2}{8}\Bigr)^{3/4} \frac{Y_e^{5/4}}{\mu^{3/4}}\, ,
\nonumber \\
\sim m_p \Bigl(\frac{e^2}{Gm_p^2}\Bigr)^{3/2} \Bigl(\frac{\eta_e}{\eta_p}\Bigr)^{3/4}  
\Bigl(\frac{27\pi^2}{8f_g^3}\Bigr)^{3/4} \frac{Y_e^{5/4}}{\mu^{3/4}}  \, ,
\nonumber \\
\propto  m_p \Bigl(\frac{\alpha}{\alpha_g}\Bigr)^{3/2} \Bigl(\frac{\eta_e}{\eta_p}\Bigr)^{3/4}\, ,
\nonumber \\
\propto  m_{pl} \eta^2_p \alpha^{3/2} \Bigl(\frac{\eta_e}{\eta_p}\Bigr)^{3/4}\, .
\end{eqnarray}
If we now set $f_g \sim 5$, and use reasonable values for $Y_e$ and $\mu$, we
finally obtain a value of $\sim$0.1 M$_{\odot}$ for $M_s$.  

Importantly, we have tethered the minimum main sequence mass to the fundamental constants
$G$, $e$, $m_e$, and $m_p$, but in fact $c$ and $\hbar$ have cancelled!
$M_s$ depends "only" on $m_p$ and the ratio $\alpha/\alpha_g$ in a duel between
electromagnetism and gravity.  The last expression for $M_s$ in eq. (\ref{mgamow})
reveals something else. It is the same as the expression in eq. (\ref{rock}) for
the mass of the largest rocky planet ($m_p \bigl(\frac{\alpha}{\alpha_g}\bigr)^{3/2}$),
but with an additional factor of $(\eta_e/\eta_p)^{3/4}$.  This factor is $\sim$300 and is, 
as one would expect, much larger than one.  As a result, we find that, given our simplifying 
assumptions and specific values for $Y_e$ and $\mu$, $M_{s}/M_{rock}$ depends only on $(\eta_e/\eta_p)^{3/4}$. 
For measured values of $m_p$ and $m_e$, $M_{s}/M_{rock}$ is then $\sim$100; 
the mass of the lowest mass star exceeds that of the most massive rocky planet. 
If we now scale $M_s$ to the detailed theoretical value of $M_{rock}$, we obtain a value 
for $M_s$ that is within a factor of 2-3 of the correct value. This quantitative correspondence 
is a bit better than might have been expected, but is heartening and illuminating nevertheless.

\section{Characteristic Mass of a Galaxy}

For stars and planets, we were not concerned with whether they could be
formed in the Universe (Nature seems to have been quite fecund, in any
case), but with the masses (say, maximum and minimum) that circumscribed
and constrained their existence. However, because of the character of
galaxies (as accummulations of stars, dark matter, and gas), the mass we
derive for them here is that for the typical galaxy in the context of
galaxy formation, and we have found this approach to be the most
productive, informative, and useful.

First, we ask: Is it obvious on simple physical grounds that a galaxy mass
will be much greater than a stellar mass, i.e. that a galaxy will contain
many stars? The distribution of the stellar masses of galaxies
($M_{star,gal}$) is empirically known to satisfy the so-called Schechter
form with probabilities distributed as: 

\begin{equation}
\label{press}
dP \propto (M_{star,gal}/M^*)^{-{\alpha_p}} e^{- M_{star,gal}/M^*} d (M_{star,gal}/M^*) \, ,
\end{equation}
with $\alpha_p$ typically having the value of $\sim$1.2 and $M^*$ a constant with units of
mass. Given this form, with a weak power law at low masses and an exponential cutoff at
high masses, there is a characteristic mass for galaxies, <$M_{star,gal}$>, the value being
somewhat greater than $M^*$, or roughly 10$^{11}$ M$_{\odot}$; galaxies much more massive than this
are exponentially rare. At the other extreme, there is very little mass bound up in
galaxies having masses much less than 10$^7$ M$_{\odot}$. The observed range of masses seems to
be set by fundamental physics in that it does not appear to be very dependent on the
epoch of galaxy formation or the environment in which the galaxies are formed (e.g., in clusters, groups, or the field).

The theory of galaxy formation is by now fairly well developed, with {\it ab initio}
hydrodynamic computations based on the standard cosmological model providing
reasonably good fits to the formation epochs, masses, sizes and spatial distributions of
galaxies, though this theory still provides rather poor representations of their detailed 
interior structures. For a recent review of some of the outstanding problems see Ostriker and Naab \cite{naab}. In the
standard $\Lambda$CDM cosmological model the mean density of matter at high redshifts is
slightly less than the critical density for bound objects to form. But there is a spectrum of
perturbations such that those which are several $\sigma$ more dense than average are
gravitationally bound and will collapse, with the dark matter and the baryons forming
self-gravitating lumps of radius $R_{halo}$, determined by the requirement that the density of the
self-gravitating system formed in the collapse is several hundred times the mean density
of the universe at the time of the collapse \cite{white}. The temperature of
the gas (absent cooling) will be the virial temperature ($C^2 \sim GM_{tot}/R_{halo}$, 
where $C$ is the speed of sound). If the gas can cool via radiative processes 
given its temperature and density, it will further collapse to the
center of the dark matter halo within which it had been embedded and will form a galaxy,
some fraction of the gaseous mass being formed into stars and a comparable fraction
ejected by "feedback" processes subsequent to star formation. 

The lower bound for normal galaxy masses is not well understood,
but is thought to be regulated by mechanical energy input
processes such as stellar winds and supernovae, all primarily
driven by the most massive ($\ge 20$ M$_{\odot}$) stars comprising
approximately one-sixth of the total stellar mass. One can easily show 
that roughly 10$^{51}$ ergs in mechanical energy input per star is
able to drive winds from star-forming galaxies with velocities 
of $\sim$300 km s$^{-1}$, and Steidel et al. \cite{steidel} have observed 
such winds to be common. Since velocities of this magnitude are comparable 
to the gravitational escape velocities for systems less massive
than our Milky Way, galaxy formation becomes increasingly
inefficient for low-mass systems and essentially ceases when
the sound speed of gas photo-heated by the ultraviolet radiation from massive
stars (10 $\rightarrow$ 20 km s$^{-1}$) approaches the escape velocity of low-mass 
dark matter halos. The result is a lower bound for normal galaxies and, 
in fact, the escape velocity from galaxies near the observed lower
bound is of order 30 km s$^{-1}$ \cite{white}.

The physical argument for the upper bound and the typical mass is somewhat more complex.
A collapsing proto-galactic clump has an evolving density ($\rho$)
and temperature ($T$).  Its collapse timescale ($t_f$) is set by the free-fall time
due to gravitation and is proportional to $1/\sqrt{G\rho}$.  As the clump
collapses and the temperature and density rise, the gas radiates photons. Its associated characteristic
cooling time ($t_c$) is the ratio of the internal energy density to the cooling rate.
Since the cooling rate per gram (see eq. \ref{cooling} below) scales as density to a higher power than the
free-fall rate, only the most overdense perturbations can cool on a time comparable to the
free-fall time. Those for which $t_c >> t_f$ will never form stars on either the free-fall time
or the only somewhat (factor of twenty) longer Hubble time. Moreover, since the highest-density 
regions are exponentially rare, perturbations having $t_c << t_f$ are uncommon. Thus, the condition, 
$t_f \sim t_c$, sets a natural and preferred scale for galaxy formation. How does this scale compare 
with the Jeans mass ($M_{Jeans}$), the mass for which gravitational and thermal 
energies are in balance and which is proportional to $T^{3/2}/\rho^{1/2}$?  In principle, the $t_f \sim t_c$ condition will
imply a relationship between $T$ and $\rho$ that might yield $M_{star,gal}$s and $M_{Jeans}$ that
are functions of $\rho$ or $T$ and, hence, may not be universal.  A wide range of
values for $M_{gal}$ would vitiate the concept of a preferred mass scale.

However, here Nature comes to the rescue.
The cooling rate (energy per volume per time) of an ideal gas
of hydrogen can be approximated (following Silk \cite{silk}, Rees \& Ostriker
\cite{rees_ostriker}, and Spitzer \cite{spitzer}) by the formula:
\begin{equation}
\label{cooling}
\Lambda_C \sim (A_{bf} + A_{ff}T) \frac{\rho^2}{T^{1/2}}\, ,
\end{equation}
where $A_{bf}$ is the bound-free (recombination) rate coefficient and $A_{ff}$ is the
corresponding free-free (bremsstrahlung) rate coefficient.  For the exploratory purposes
of this study, these coefficients are:
\begin{eqnarray}
\label{abf}
A_{ff} \sim \frac{2^{9/2}\pi^{1/2}}{3^{3/2}} \frac{e^4\alpha k_B^{1/2}}{m_e^{3/2}m_p^2 c^2}  \, ,
\nonumber \\
A_{bf} \sim \alpha^2 \frac{m_e c^2}{k_B}A_{ff}\, .
\end{eqnarray}
In eqs. (\ref{cooling}) and (\ref{abf}), $m_e$, $e$, and $\alpha$ appear due
to the importance of electromagnetic radiation processes.  As eq. (\ref{cooling})
suggests, free-free cooling exceeds bound-free cooling at high temperatures.
Equation (\ref{abf}) indicates that the crossover temperature, below which
recombination cooling predominates, is $\sim \alpha^2 \frac{m_e c^2}{k_B}$,
which, using measured numbers, is $\sim$$3 \times 10^5$ K.  The
temperatures of relevance during the incipient stages of galaxy formation
are not much larger than this, so we can neglect the $A_{ff}$ term in eq. (\ref{cooling})
and find that $\Lambda_C \propto \frac{\rho^2}{T^{1/2}}$. For an ideal gas,
the internal energy density is $\frac{3/2\rho k_B T}{\mu {m_p}}$.  Therefore,
$t_c \sim \frac{\rho k_B T}{m_p \Lambda_C} \propto T^{3/2}/\rho$.
Since $t_f \propto 1/\rho^{1/2}$, $t_c/t_f$ is proportional to $T^{3/2}/\rho^{1/2}$ $-$
this is proportional to $M_{Jeans}$, the Jeans mass!
So, we find that the $t_c/t_f \sim 1$ condition filters out a specific mass.
What is its value? From $t_c/t_f \sim 1$
and the proper expression for $M_{Jeans}$ ($= M_{gal}$), we derive:
\begin{eqnarray}
\label{mgal}
M_{gal} \sim m_p \frac{\alpha^5}{\alpha_g^2} \Bigl(\frac{\eta_e}{\eta_p}\Bigr)^{1/2}  \, ,
\nonumber \\
\sim M_{Ch} \eta_p \alpha^5 \Bigl(\frac{\eta_e}{\eta_p}\Bigr)^{1/2} \, ,
\nonumber \\
\sim m_{pl} \eta^3_p  \alpha^5 \Bigl(\frac{\eta_e}{\eta_p}\Bigr)^{1/2} \, .
\end{eqnarray}
Note that, with eq. (\ref{mgal}) we have obtained $M_{gal}$ not only in terms of 
$m_{pl}$, $\eta_p$, $\eta_e$ and $\alpha$, but should we wish to so express it, in terms 
of the familiar constants $G$, $\hbar$, and $c$ (as well as $\eta_e$, $\eta_p$, and $\alpha$).  
Note also that $\eta_p$ is a very large number and more than compensates for the smallness of $\alpha^5$.
The appearance of $m_e$, $c$, and $\alpha$ is a natural consequence
of cooling's dependence on electromagnetic processes.  As an indication of the
importance of quantum mechanics in determining galaxy characteristics, $\hbar$ does not cancel.

For measured values of the fundamental constants, and retaining
the prefactor dropped in eq. (\ref{mgal}) ($\frac{2^{5/2}\pi^{7/2}}{27}$), but retained in Silk (1977),
we find $M_{gal} \sim 10^{11}$ M$_{\odot}$, reassuringly close to the characteristic mass in stars ($M^*$)
of the "average" $L^*$ galaxy in our Universe.
Moreover, one can derive a Jeans length and, hence, a length scale for this average galaxy.
It is:
\begin{equation}
\label{rgal}
R_{gal} \sim \frac{\hbar}{m_e c} \alpha^3 \eta^2_p \Bigl(\frac{\eta_e}{\eta_p}\Bigr)^{1/2}
\nonumber\\
\sim R_{pl} \alpha^3 \Bigl(\frac{\eta_e}{\eta_p}\Bigr)^{3/2}\, ,
\end{equation}
an expression that scales with the Compton wavelength of the electron. Plugging in measured numbers
gives $R_{gal} \sim 50$ kiloparsecs, a number well within reason in our Universe (and, in fact,
for our own Milky Way).

While it is reassuring that the critical mass that appears from our dimensional analysis
corresponds well to the upper mass range of normal galaxies, we are left with two
questions. First, there exist galaxies which are up to $\sim$10 times greater in mass than $M^*$; how do
these form? And, second, what happens to dark matter lumps that are much more massive
than this critical mass and are dense enough to collapse? The answer to the first question
is becoming observationally clear. All of these supergiant galaxies are "BCGs", brightest
cluster galaxies. We now know that they form at early times, reach a mass comparable to
$M^*$, cease star formation but keep on growing in mass (by roughly a factor of 2$-$3) and in
size (by roughly a factor of 4$-$8). The process by which this happens is 
"Galactic cannibalism" \cite{hausman} by which gravitationally induced
dynamical friction causes the inspiral of satellite galaxies to merge with the central
galaxy. Thus, the "excessive" mass of BCGs is caused by a distinct process of mass growth.
With regard to the second question, the answer again lies in observations of
groups and clusters. If these giant, self-gravitating units ("dark matter halos") have
total masses far above ($\Omega_{matter}/\Omega_{baryon}$) $M^*$, 
i.e. greater than 10$^{12}$ M$_{\odot}$, then they host not one 
giant mass galaxy, but rather a distribution of galaxy masses, the distribution given by
eq. \ref{press}\footnote{$\Omega$ is the ratio of the density of a mass-energy component of the
Universe to the "critical" total density.  Here, $\Omega_{matter}/\Omega_{baryon}$ is the ratio
of the total matter density (dark matter plus regular matter) to the density
of regular matter ("baryons").}. Thus, we find it natural that most of the mass in the universe is in stellar
systems containing roughly 10$^{11}$ stars, each with mass between the limits $M_s$ and $M_S$,
which both scale with the Chandrasekhar mass.

\section{Conclusion}
\label{conclusion}

We now recapitulate in succinct form most of the masses discussed
in this paper. First, we express our results in the most fundamental units:

\begin{eqnarray}
\label{eq.conclusion2}
M_{rock} &\sim& m_{pl} \eta^2_p \alpha^{3/2}  \, ,
\nonumber \\
M_s &\sim& m_{pl} \eta^2_p \alpha^{3/2} \Bigl(\frac{\eta_e}{\eta_p}\Bigr)^{3/4} \, ,
\nonumber \\
M_{S} &\sim& 50\, m_{pl} \eta^2_p \, ,
\nonumber \\
M_{Ch} &\sim& m_{pl} \eta^2_p \, ,
\nonumber \\
M_{NS} &\sim&   m_{pl} \eta^2_p  \Bigl(\frac{\eta_{\pi}}{2\beta_n{\eta_{p}}}\Bigr)^{3/2} \, ,
\nonumber \\
M_{ns} &\sim& m_{pl} \eta^2_p \Bigl(\frac{\eta_{p}}{\eta_{\pi}}\Bigr)^{3} \, ,
\nonumber \\
M_{gal} &\sim& m_{pl} \eta^3_p {\alpha^5} \Bigl(\frac{\eta_e}{\eta_p}\Bigr)^{1/2}\, .
\end{eqnarray}
Equation (\ref{eq.conclusion2}) reduces the maximum mass of a rocky planet ($M_{rock}$), 
the minimum mass of a star ($M_s$), the maximum mass of a star ($M_S$), the maximum mass of a white dwarf ($M_{Ch}$), 
the maximum mass of a neutron star ($M_{max}$), the minimum mass of a neutron 
star ($M_{ns}$), and the characteristic mass of a galaxy ($M_{gal}$) 
to only five simple quantities.  
and makes clear a natural mass hierarchy related only to $m_p$, particle mass ratios, and
the "$\alpha$s."  We have expressed the basics of  
important astronomical objects with only five constants, modulo some dimensionless numbers 
of order unity.  Equations (\ref{eq.conclusion2}) summarize the interrelationships 
imposed by physics between disparate realms of the Cosmos.  

Alternatively, it is instructive to put these same relations into a somewhat more familiar
form. We drop all dimensionless constants of order unity and summarize the relations derived for
the masses of astronomical bodies in units of the Chandrasekhar mass (which is close to a
solar mass), the three particle masses ($m_p$, $m_{\pi}$, $m_e$) and the Planck mass, 
$m_{pl}$, with $M_{Ch} \sim m_p (m_{pl}/ m_p)^3 \sim 1 $M$_{\odot}$.

We found that neutron stars can exist within the range

\begin{equation}
(m_{\pi}/ m_p)^3 < M/M_{Ch} < (m_p/m_{\pi})^{3/2}\, .
\nonumber\\
\end{equation}

Normal stars can exist within the range

\begin{equation}
((m_p/ m_e) \alpha^2)^{3/4} < (M/ M_{Ch}) < \quad \sim 50 \, ,
\nonumber\\
\end{equation}

and rocky planets can exist with masses

\begin{equation}
M/ M_{Ch} < \alpha^{3/2} \, ,
\nonumber\\
\end{equation}
which is comfortably smaller than the minimum mass of stars.
Finally, normal galaxies have a characteristic mass

\begin{equation}
M/ M_{Ch} \sim \alpha^5 (m_{pl}/ m_p) (m_p/ m_e)^{1/2}\, ,
\nonumber\\
\end{equation}
which is larger than the characteristic mass of both normal low mass stars and even the
most massive stars by a very large factor.

Everything astronomical is indeed connected, and that the essence of an object can be reduced to a few central 
quantities is one of the amazing consequences of the unifying character of physical law.  
Indeed, in this exercise we focussed on stars, planets, and galaxies, 
avoided complexity, eschewed any hint that emergent phenomena might be of 
fundamental import, and ignored topics such as life and the ubiquitous complexity that 
clutters most experience. 
%
%
Rather, our goal here has been to understand and articulate the simple connections inherent in 
the universal operation of a small number of physical principles 
and fundamental constants, and to identify the ties 
between seemingly unrelated, but key, astronomical entities. 
We hope we have conveyed to the reader that not only are these connections
knowable and quantifiable, but that they are both simple and profound.  
%

\begin{acknowledgments}
The authors thank Jeremy Goodman and Neta Bahcall for their insightful comments
on an earlier draft of this manuscript and Paul Langacker for stimulating conversations.
\end{acknowledgments}

\end{article}

\end{document}